\documentclass[sigconf]{acmart}

\AtBeginDocument{%
  \providecommand\BibTeX{{%
    \normalfont B\kern-0.5em{\scshape i\kern-0.25em b}\kern-0.8em\TeX}}}

\copyrightyear{}
\acmYear{}
\setcopyright{none}
\acmConference[EASE 2021]{Evaluation and Assessment in Software Engineering}{June 21--23, 2021}{Trondheim, Norway}
\acmBooktitle{}
\acmPrice{}
\acmDOI{10.1145/3463274.3463361}
\acmISBN{}

\begin{document}

\title{Benchmarking as Empirical Standard in Software Engineering Research}

\author{Wilhelm Hasselbring}
\email{hasselbring@email.uni-kiel.de}
\orcid{0000-0001-6625-4335}
\affiliation{%
  \institution{Kiel University}
  \streetaddress{Christian-Albrechts-Platz 4}
  \city{Kiel}
  \state{}
  \country{Germany}
  \postcode{24118}
}

\renewcommand{\shortauthors}{Hasselbring}

\begin{abstract}
In empirical software engineering, benchmarks can be used for comparing different methods, techniques and tools.
However, the recent ACM SIGSOFT Empirical Standards for Software Engineering Research do not include an explicit checklist for benchmarking. 
In this paper, we discuss benchmarks for software performance and scalability evaluation as example research areas in
software engineering, relate benchmarks to some other empirical research methods, and discuss the requirements on benchmarks that may constitute the basis for a checklist of a benchmarking standard for empirical software engineering research.
\end{abstract}

\ccsdesc[500]{Software and its engineering}

\keywords{benchmarking, empirical software engineering, empirical standards}

\maketitle

\section{Introduction}

The recent ACM SIGSOFT Empirical Standards for Software Engineering Research~\cite{ralph2021acm} are meant to give reviewers guidelines to evaluate manuscripts against expectations determined by a scientific community. Empirical standards describe these community expectations. If publication venues adopt these standards, authors know the expectations in advance and can follow the essential criteria laid out by the standard. Reviewers then check submitted papers against the specific criteria in the relevant standards. These standards may also help to educate the next generations of software engineering researchers.

The ACM SIGSOFT Empirical Standards provide a catalog of such standards, starting with the General Standard that applies to all empirical research. Next, there is a set of methodology-specific standards such as experiments, questionnaire surveys and case studies. Several  \textit{supplements} for cross-cutting concerns like information visualization and sampling complement the catalog. The standards catalog is modular to reduce duplication, so most research projects will use multiple standards.
Each empirical standard is essentially a one-page checklist of specific criteria that can be used by authors
to conduct and report research, and by reviewers to evaluate manuscripts.

In empirical software engineering, benchmarks can be used for comparing different methods, techniques and tools. Tichy~\cite{tichy_ubiquity_2014} summarizes the benefits for benchmarks for software research as follows:
\begin{quote}
``Software research could benefit tremendously from benchmarks. Benchmarks can be tested repeatedly and quickly without requiring human subjects. They help weed out poor techniques quickly and direct attention to the successful ones. There is a certain upfront cost for constructing the benchmark, but that effort could be shared among many researchers.''
\end{quote}
Benchmarking should be added to the arsenal of empirical methods in order to speed up progress~\cite{Tichy1998,tichy_ubiquity_2014}.
The creation and widespread use of a benchmark within a research area is frequently accompanied by rapid technical progress and community building~\cite{Sim2003}. The Darpa Grand Challenge for self-driving cars represents an example for rapid technical progress inspired by benchmarks~\cite{tichy_ubiquity_2014}. The existence of a benchmark is indicative of the maturity of a scientific
discipline~\cite{Sim2003}.

Benchmarks are used to compare different platforms, methods, tools, or techniques. They define standardized measurements to provide repeatable, objective, and comparable results~\cite{kounev2020systems}. 
In computer science, benchmarks are used to compare, for instance, the performance of database management systems~\cite{Gray93},  information retrieval algorithms~\cite{Nanni2017} and cloud services~\cite{bermbach2017cloud}.

For (quantitative) simulations, for instance, there exists a checklist in the Empirical Standards.\footnote{\url{https://github.com/acmsigsoft/EmpiricalStandards/blob/master/docs/QuantitativeSimulation.md}}
So far, there exists no such checklist for benchmarks in the Empirical Standards.
However, benchmarking -- similar to simulation -- is relevant for evaluating engineering research, which is research that invents and evaluates technological artifacts~\cite{Empirical2021}. This is already mentioned in the following two quotes of the Engineering Research Standard:\footnote{\url{https://github.com/acmsigsoft/EmpiricalStandards/blob/master/docs/EngineeringResearch.md}}
\begin{itemize}
	\item ``[\dots] empirically compares the artifact to one or more state-of-the-art benchmarks''
	\item ``Antipatterns: [\dots] evaluation consists \textit{only} of quantitative performance data that is not compared to established benchmarks or alternative solutions''
\end{itemize}
Thus, benchmarking is already mentioned to assess new results of engineering research.
However, benchmarking is not, as yet, described as an empirical standards on its own.

Section~\ref{s-performance} discusses benchmarks for software performance and scalability evaluation as example research areas in software engineering. We relate benchmarks to some other empirical methods in Section~\ref{s-related}. Section~\ref{s-requirements} discusses the requirements on benchmarks before Section~\ref{s-conclusion} concludes the paper.

\section{Benchmarks for Software Performance and Scalability Evaluation}\label{s-performance}

Performance benchmarks are part of the measurement-based approaches in the field of Software Performance Engineering~\cite{smith2002performance}.
The employment of performance benchmarks has contributed to improve generations of systems~\cite{Vieira2012}.

In the following subsections, we briefly present a few exemplar benchmarks for software performance and scalability evaluation to illustrate benchmarking in software engineering research. Section~\ref{s-Teastore} presents the TeaStore benchmark that provides an example microservice-based software application together with synthetic workloads to execute the benchmarks. Section~\ref{s-MooBench} presents the MooBench benchmark to measure the performance overhead of monitoring frameworks, with an emphasis on integrating regression benchmarking into continuous integration pipelines.
The Theodolite scalability benchmark for distributed stream processing engines is presented in Section~\ref{s-Theodolite}.

\subsection{The TeaStore Benchmark for Performance and Scalability Benchmarking}\label{s-Teastore}

The TeaStore is an online store for tea and tea related utilities~\cite{vonKistowski2018}.
The TeaStore benchmark application has been used as a distributed system for evaluating and extracting software performance models, for testing single and multi-tier auto-scalers, and for software energy-efficiency analysis and management.

\begin{figure}[htb]
 \centering
\includegraphics[width=\linewidth]{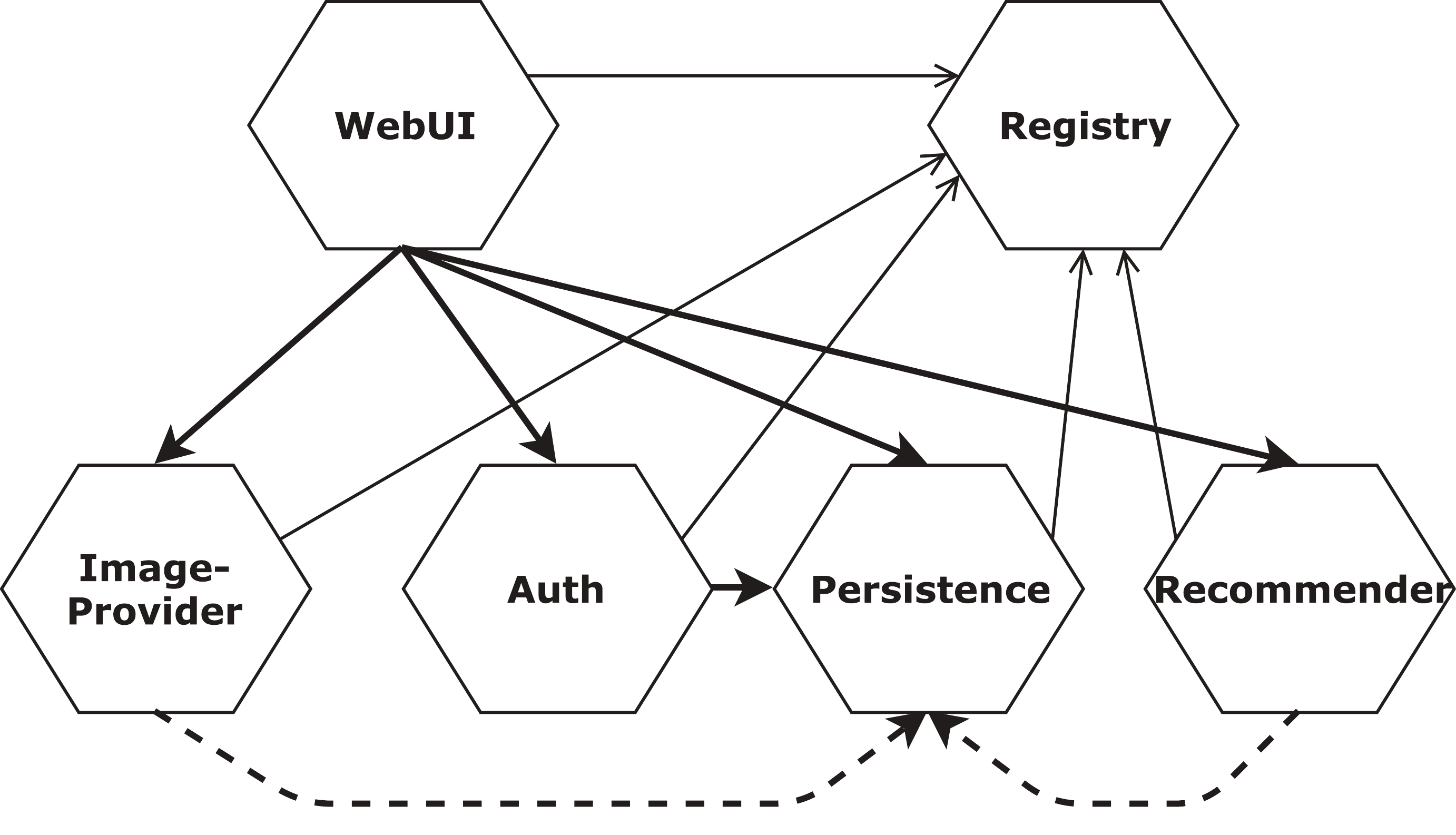}
  \caption{\label{f-TeaStoreArchitecture}TeaStore Architecture~\cite{vonKistowski2018}.}
	\Description{TeaStore Architecture}
\end{figure}

The TeaStore software architecture consists of five distinct services and a Registry service as shown in Figure~\ref{f-TeaStoreArchitecture}. All services communicate with the Registry. Additionally, the WebUI service issues calls to the Image-Provider, Authentication, Persistence and Recommender services.
The TeaStore uses a client-side load balancer to allow replication of instances of one service type.

As the TeaStore is a benchmarking application, it is open source and available to instrumentation using available monitoring solutions. Pre-instrumented Docker images for each service include the Kieker monitoring framework~\cite{KiekerICPE2012,Kieker2020}.

\begin{figure*}[htb]
 \centering
\includegraphics[width=.9\linewidth]{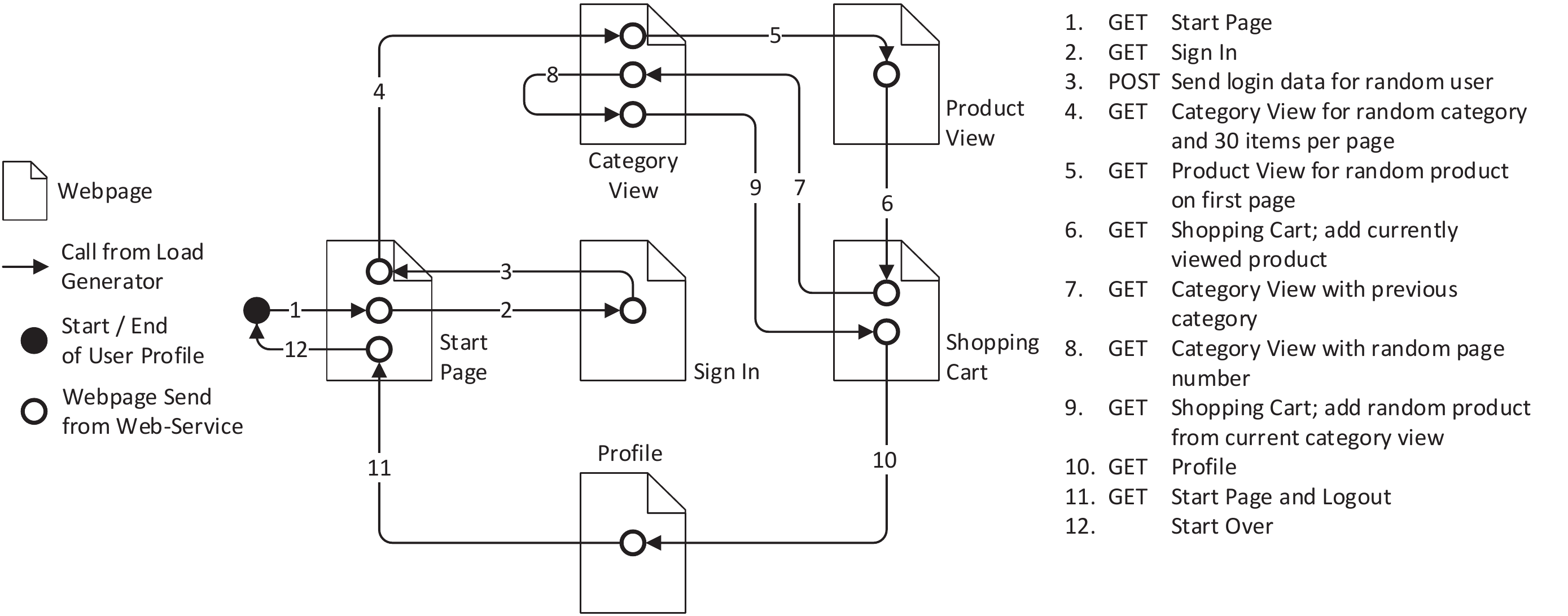}
  \caption{\label{f-TeaStoreProfile}Browse user profile~\cite{vonKistowski2018}.}
	\Description{Browse user profile}
\end{figure*}

\begin{figure}[htb]
 \centering
\includegraphics[width=\linewidth]{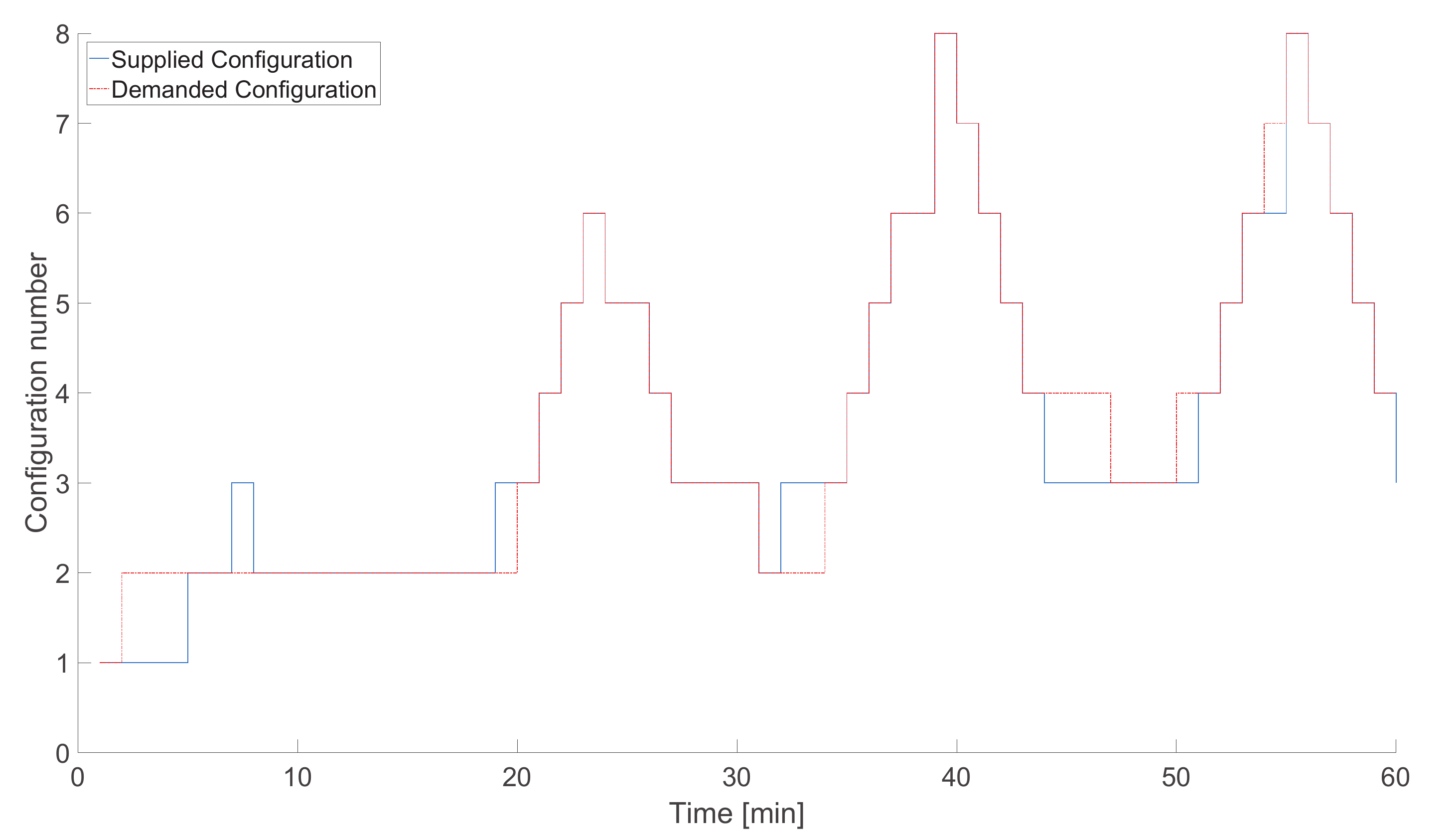}
  \caption{\label{f-TeaStoreFIFA}Scaling behavior for the FIFA trace~\cite{vonKistowski2018}.}
	\Description{Scaling behavior for the FIFA trace}
\end{figure}

\begin{figure}[htb]
 \centering
\includegraphics[width=\linewidth]{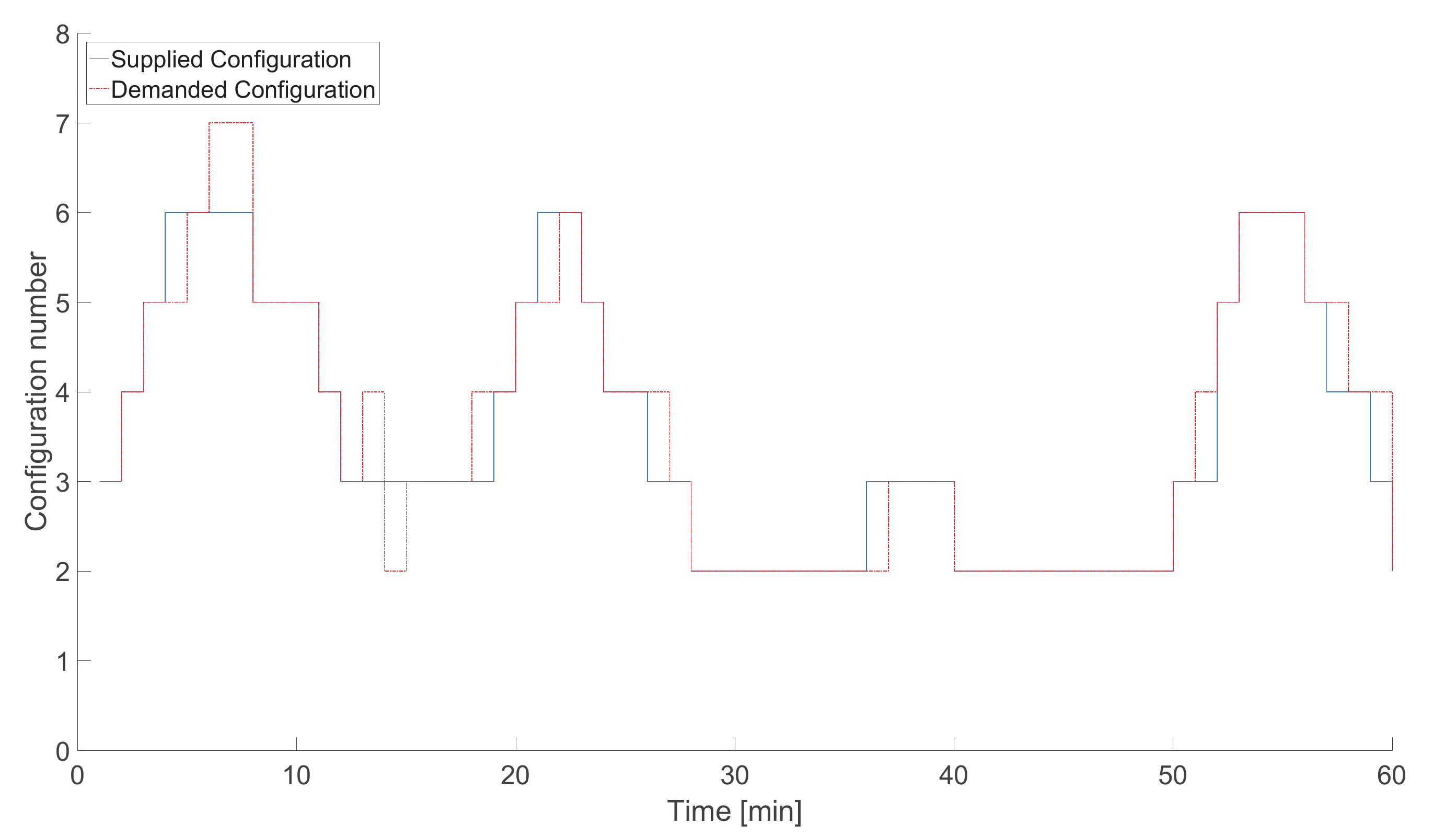}
  \caption{\label{f-TeaStoreBibSonomy}Scaling behavior for the BibSonomy trace~\cite{vonKistowski2018}.}
	\Description{Scaling behavior for the BibSonomy trace}
\end{figure}

Besides the TeaStore application, the benchmark provides synthetic user profiles for automated load testing.
Figure~\ref{f-TeaStoreProfile} shows the Browse user profile. Users log in, browse the store for products, add these products to the shopping cart and then log out. The number of users is chosen depending on the maximum load.
Besides such artificial user profiles, TeaStore employs synthetic workloads that are derived from two real-world traces (FIFA World Cup 1998~\cite{Arlitt2000} and BibSonomy~\cite{Benz2010}). The TeaStore application employs an auto-scaler to automatically scale the store at run-time as the load intensity varies. 
The scaling behavior on both the FIFA and BibSonomy traces are shown in Figure~\ref{f-TeaStoreFIFA} and in Figure~\ref{f-TeaStoreBibSonomy}. Both figures are structured as follows: The horizontal axis shows the experiment time in minutes; the vertical axis represents the current number of scaling units.

In Figure~\ref{f-TeaStoreFIFA}, for example, the system is in an under-provisioned state in the entire interval between minute 2 and 5. Overall, the under-provisioning and over-provisioning time-shares indicate good scaling behavior in this experiment~\cite{vonKistowski2018}.

\begin{figure}[htbp]
\centerline{\includegraphics[width=\linewidth]{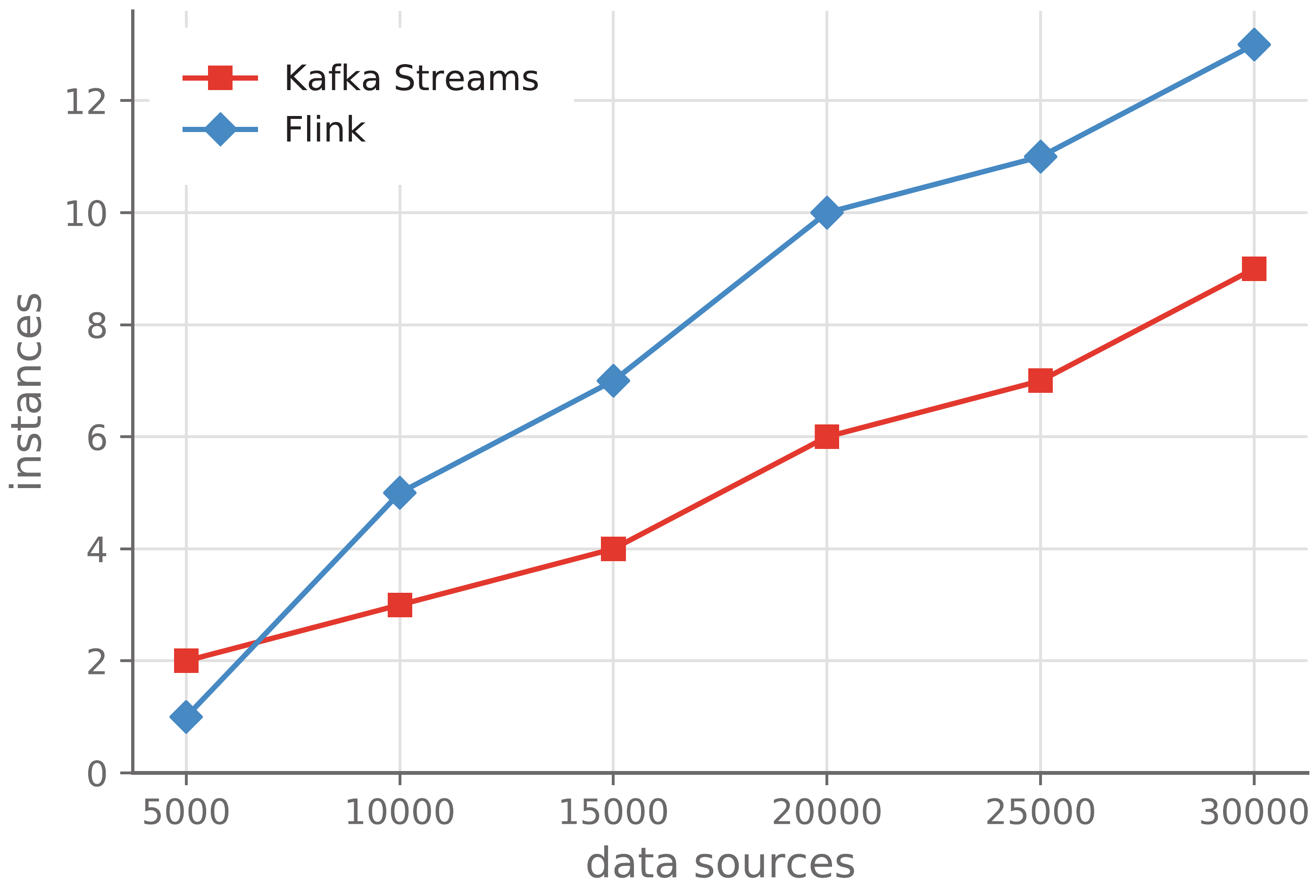}}
  \caption{\label{f-Theodolite}Comparison of Kafka Streams and Flink via Theodolite's benchmark for aggregating time attributes \cite{BDR2021}.}
	\Description{Comparison of Kafka Streams and Flink via Theodolite's benchmark for aggregating time attributes}
\end{figure}

\subsection{The MooBench Monitoring Benchmark}\label{s-MooBench} 

The MooBench~\cite{MSEPT2012} micro-benchmark has been developed to quantify the overhead for application-level monitoring frameworks under controlled and repeatable conditions.
MooBench has also been used by other researchers for replicable performance experiments comparing monitoring frameworks~\cite{Knoche2018}; thus, fostering research on monitoring frameworks.

\begin{figure*}[htb]
 \centering
\includegraphics[width=.7\linewidth]{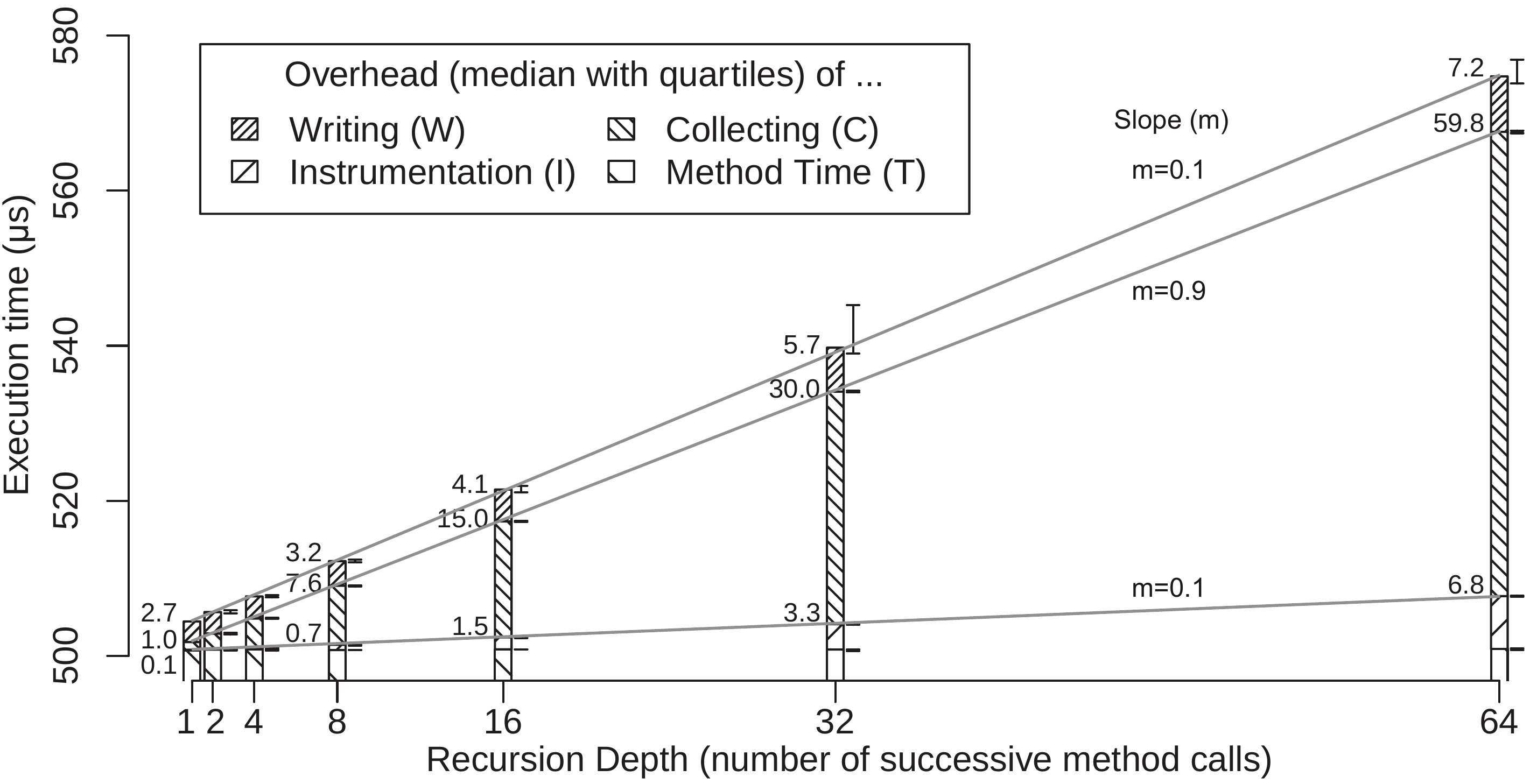}
  \caption{\label{f-MooBench}Benchmarking the performance overhead of monitoring frameworks with the MooBench benchmark \cite{MSEPT2012}.}
	\Description{Monitoring Overhead with MooBench}
\end{figure*}

Waller \&\ Hasselbring~\cite{MSEPT2012} employ the MooBench benchmark to evaluate the monitoring overhead of the Kieker~\cite{KiekerICPE2012,Kieker2020} monitoring framework and to measure the influence of different configurations for multi-core processors in this context. Fig.~\ref{f-MooBench} shows the linear increase of the overhead with MooBench, when applied to Kieker. 

\begin{figure*}[htbp]
\centerline{\includegraphics[width=.9\linewidth]{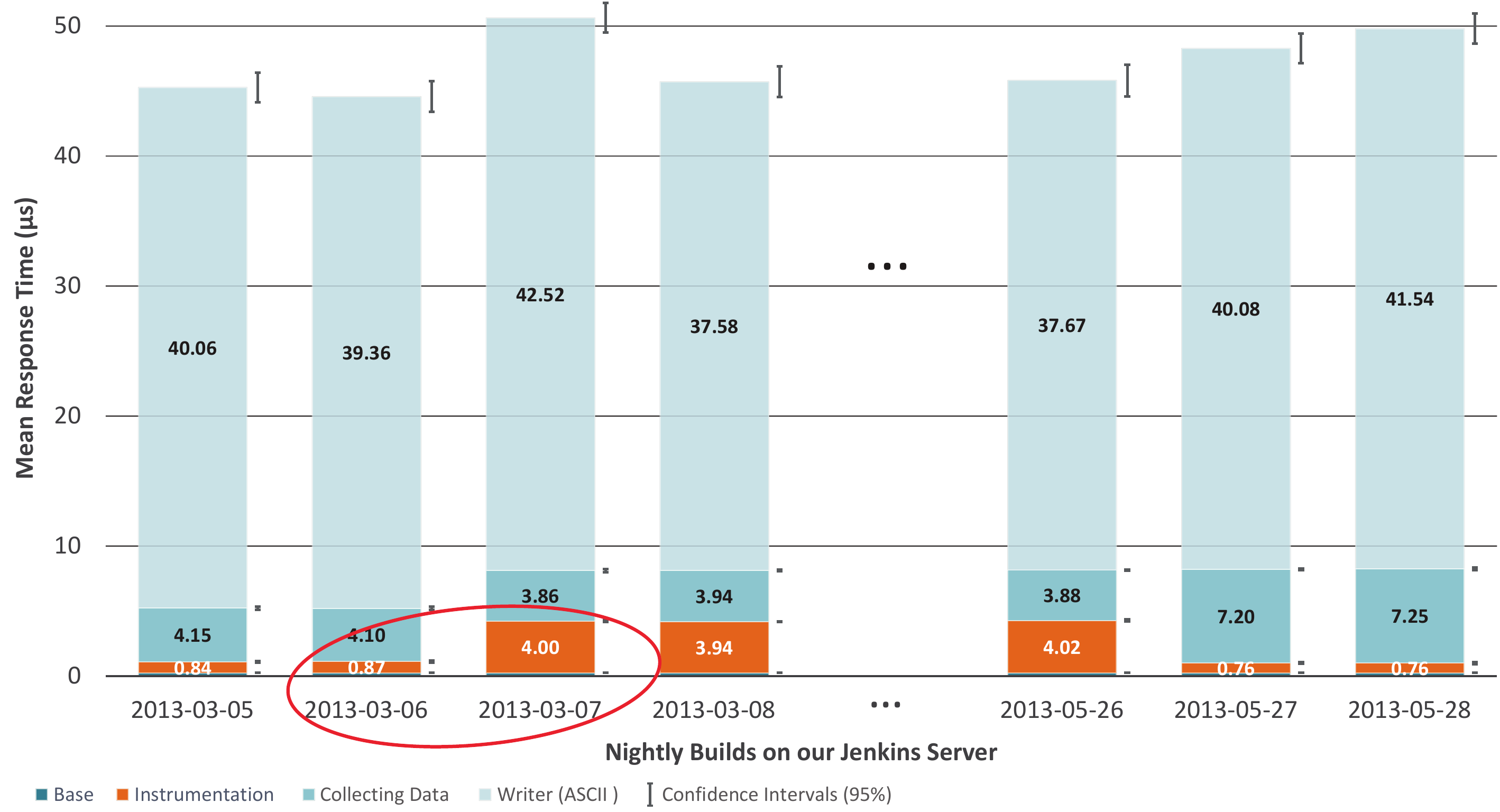}}
  \caption{\label{f-MooBenchJenkins}Scenario for detecting performance anomalies between releases via benchmarks in continuous integration~\cite{SEN2015}.}
	\Description{Scenario for detecting performance anomalies between releases via benchmarks in continuous integration}
\end{figure*}

Moobench has been integrated into the continuous integration pipeline of Kieker, allowing for automatic regression benchmarking~\cite{SEN2015}. An example visualization of a series of regression benchmark results is presented in Figure~\ref{f-MooBenchJenkins}. It shows a performance regression that happened in March 2013 with Kieker release version 1.7, and later diagnosed as a bug in the implementation of adaptive monitoring. 
The daily results of Kieker with MooBench may be consulted at \url{http://kieker-monitoring.net/performance-benchmarks/}.

\subsection{The Theodolite Scalability Benchmark for Distributed Stream Processing Engines}\label{s-Theodolite}

Scalability is usually defined as the ability of a system to continue processing an increasing workload with additional resources provided~\cite{Elasticity2013}. Whereas benchmarking the \textit{performance} of stream processing engines such as throughput or latency is heavily performed by academia and industry \cite{Karimov2018}, approaches for benchmarking their \textit{scalability} are scarce.
Theodolite~\cite{BDR2021} provides a method for benchmarking the scalability of distributed stream processing engines. With Theodolite, individual benchmarks are designed based on typical use cases for stream processing within microservices~\cite{Hasselbring2017,SA2018}. Microservice architectures aim, in particular, at scalability~\cite{ICPE2016Keynote,EMISA2019}. Theodolite supports evaluating scalability independently along different dimensions of increasing workloads.
As an example, Fig.~\ref{f-Theodolite} compares the results for Kafka Streams and Flink via Theodolite's benchmark for aggregating time attributes.

\begin{figure*}[htbp]
	\centering%
	\includegraphics[width=\textwidth]{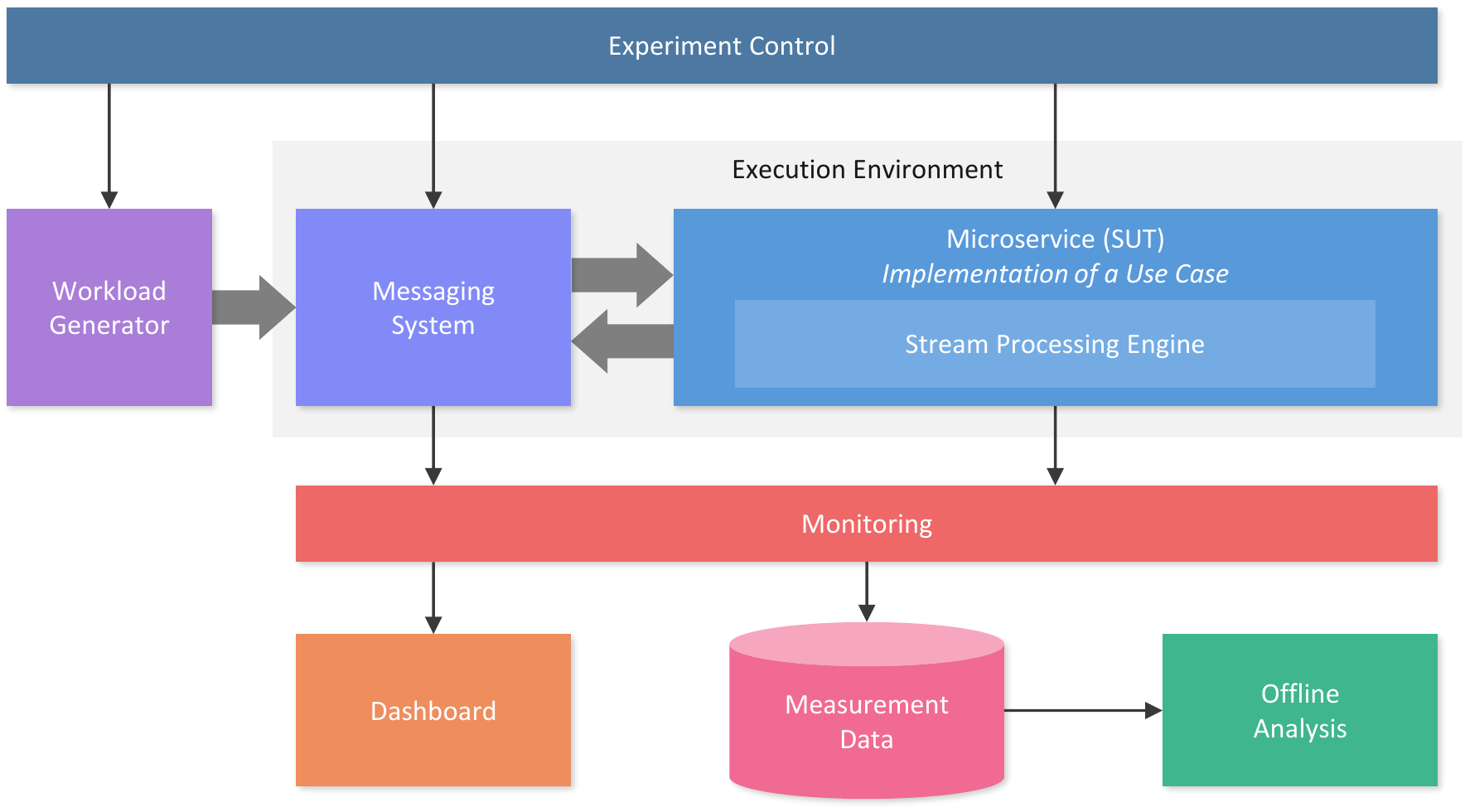}%
	\caption{\label{fig:execution-architecture}The Theodolite framework architecture for executing scalability benchmarks \cite{BDR2021}.}
	\Description{The Theodolite framework architecture for executing scalability benchmarks}
\end{figure*}%

A benchmark such as Theodolite does not only consist of benchmarking data, but it provides a software framework for executing the benchmarks. 
Figure~\ref{fig:execution-architecture} depicts the Theodolite framework architecture for executing scalability benchmarks. It consists of the following components:
\begin{description}
	\item[Experiment Control]
The central experiment control is started at the beginning of each scalability benchmark and runs throughout its entire execution. 
For each experiment, it starts and configures the workload generator component to generate the current workload of the tested dimension. Further, it starts and replicates the SUT (system under test) according to the evaluated number of instances.
After each experiment, this component resets the messaging system, ensuring no queued data can be accessed by the following subexperiment.

	\item[Workload Generator]
This component generates a configurable constant workload of a configurable workload dimension. It fulfills the function of a data source in a big data streaming system, such as an IoT device or another microservice.
Since different use cases require different data input formats, Theodolite allows for individual workload generators per use case. However, individual workload generators can share large parts of their implementations.

	\item[Messaging System]
In event-driven, microservice-based architectures, individual services usually communicate with each other via a dedicated messaging system. The Theodolite benchmarking architecture therefore contains such a system, serving as a message queue between workload generator and stream processing engine and as a sink for processed data.
State-of-the-art messaging systems already partition the data for the stream processing engine and are, thus, likely to have high impact on the engine's scalability. They provide plenty of configuration options, making it reasonable to benchmark different configurations against each other.

	\item[Microservice (SUT)]
This component acts as a microservice that applies stream processing and, thus, is the actual SUT. This microservice fulfills a specific use case.
An implementation of this microservice uses a certain stream processing engine along with a certain configuration, which should be benchmarked.
The stream processing engine receives all data to be processed from the messaging system and, optionally, writes processing results back to it.

	\item[Monitoring]
The monitoring component collects runtime information from both the messaging system and the stream processing engine. This includes data to be displayed by the dashboard and data required to actually measure the scalability of the SUT.

	\item[Dashboard]
Our Theodolite architecture contains a dashboard for observation of benchmark executions. It visualizes monitored runtime information of the execution environment, the messaging system, and the SUT. Thus, it allows to verify the experimental setup (e.g., number of deployed instances and number of generated messages per seconds).

	\item[Offline Analysis]
Based on the raw monitoring data, a dedicated component evaluates the scalability of the SUT by computing the required metrics. This component is executed offline after completing all experiments. Since Theodolite stores monitoring data persistently, you can repeat all computations at any time without re-executing the underlying experiments.
\end{description}

The Theodolite benchmarking framework can be configured by the following parameters:
\begin{enumerate}
	\item An implementation of the use case that should be benchmarked
	\item Configurations for the SUT including messaging system and execution environment
	\item The workload dimension for scalability benchmarking
	\item A workload generator generating workloads along the configured dimensions
	\item A list of workloads for the configured dimension to be tested
	\item A list of numbers of instances to be tested
\end{enumerate}
For details refer to Henning \&\ Hasselbring~\cite{BDR2021}.

\subsection{Summary}

TeaStore, MooBench and Theodolite are example benchmarks for software performance and scalability evaluation.
So far, the Empirical Standards do not include review guidelines for such benchmarking experiments. 

\clearpage
\section{Benchmarks related to other Empirical Research Methods}\label{s-related}

Empirical research methods related to benchmarking are simulations, case studies and experiments.

\subsection{Benchmarks vs.\ Simulations}

A simulation study involves developing and using a mathematical model that imitates a real-world system's behavior.
In computational science~\cite{CiSE2018}, such as ocean science~\cite{Johanson:2017,ESE2017}, we have a clear separation between the model and the real world. In computer science and software engineering this is not always the case: both, the implemented model and the real-world system are software systems.

\begin{figure*}[btp]
  \centering
  \includegraphics[width=.8\textwidth]{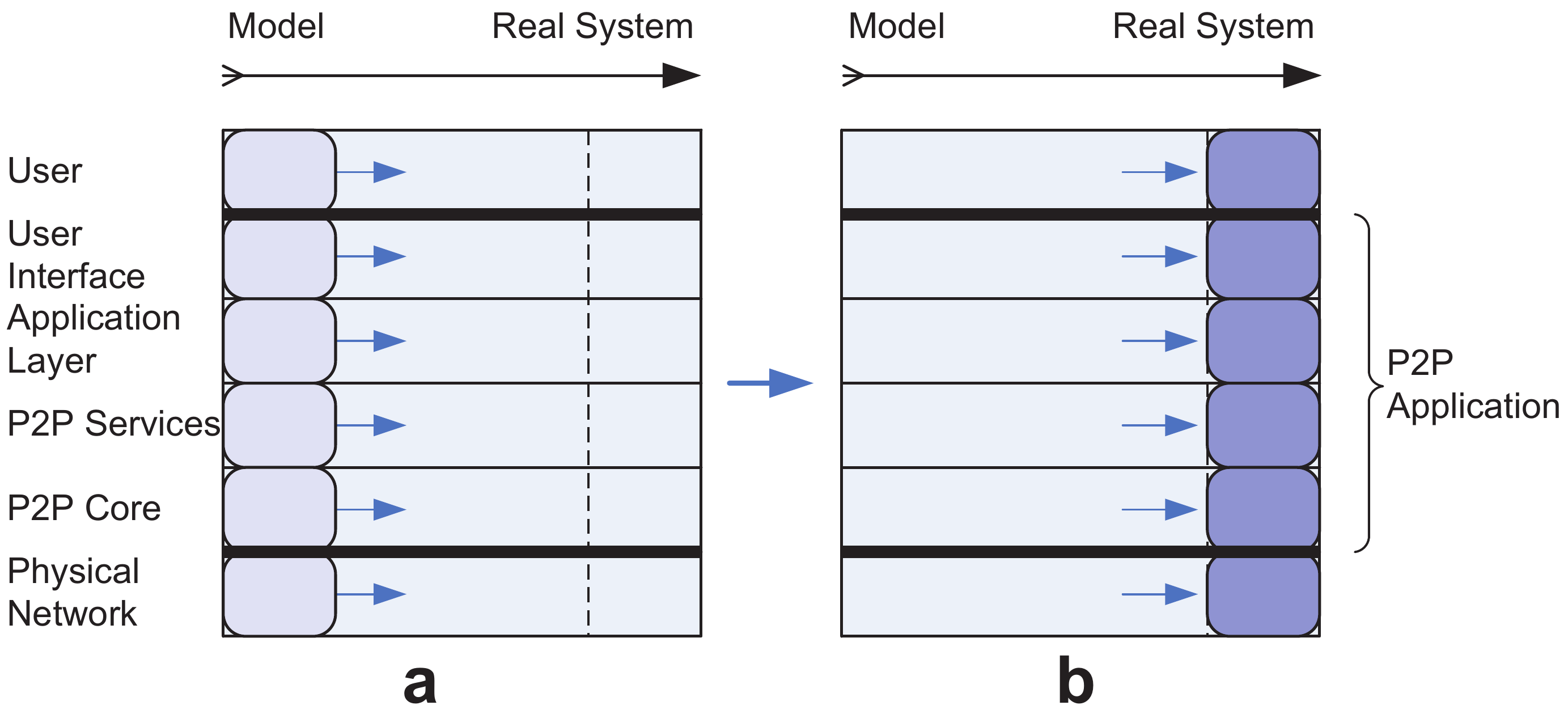}
  \caption{\label{f-RealPeer}Simulation-based development of P2P systems with RealPeer \cite{RealPeer2008}.}
	\Description{Simulation-based development of P2P systems with RealPeer}
\end{figure*}

Let us take a look at Peer-to-Peer (P2P) systems as an example domain.
In the process of developing P2P systems, simulation has proved to be an essential method for the evaluation of new P2P algorithms and system architectures. The separation between a simulation model of a P2P system and a real P2P system that operates on a real physical network hinders the transition of simulation models to real systems.
To bridge this gap, RealPeer \cite{Hildebrandt2007,RealPeer2008} introduces the idea of \textit{simulation-based development} of P2P systems~\cite{JSS2006}. With RealPeer, an initial simulation model of a P2P system is iteratively transformed into the intended real P2P system. The RealPeer framework supports a developer in modeling, simulating and ultimately developing P2P systems.

Fig.~\ref{f-RealPeer} illustrates the use of layered models for simulation-based development of P2P systems with RealPeer. Initially, for each layer of the model a developer creates a model of the corresponding aspect (left hand side). These models are incrementally refined until they correspond to the intended real P2P system at the end of the development process (right hand side). The last step (separated by a dashed line) is a special case. In this step, each element of the model is replaced by its real counterpart. 
For a full introduction to RealPeer, refer to Hildebrandt \&\ Hasselbring~\cite{RealPeer2008}.

RealPeer supports the modeling and simulation as well as the development of P2P systems.
The resulting real P2P system -- not the simulation -- could eventually become a candidate for a P2P benchmark.

Simulations and benchmarks have a lot in common, for instance, we can \textit{measure} the performance of software systems via benchmarks or \textit{assess} the performance via simulation~\cite{ECSA2011Massow}.
However, the essential difference is that simulations execute a \textit{model} of a system while benchmarks execute the \textit{actual} system (within some experiment setup).

\subsection{Benchmarks vs.\ Case Studies}

Benchmarks require specified, synthetic workloads~\cite{vanHoorn2008}, while case study workloads should originate from real-live usage.
Case studies are an empirical inquiry that investigates a contemporary phenomenon (the `case') in depth and within its real-world context. Unfortunately, the label `case study' is often misused in the software engineering literature~\cite{Wohlin2021}. For instance, illustrative examples are often called case study. But neither examples nor benchmarks are case studies according to the Empirical Standards.\footnote{\url{https://github.com/acmsigsoft/EmpiricalStandards/blob/master/docs/CaseStudy.md}}

The synthetic workloads for benchmarks may be derived from real-live usage~\cite{ValueTools2014,WESSBAS2018}, but for comparative and reproducible evaluation of the systems under test we need defined synthetic workloads.

An advantage of benchmarking over case studies is that replication is built into the method.

\subsection{Benchmarks vs.\ Controlled Experiments}

Experiments require a high level of control over all variables affecting the outcome but also provide reproducibility and easy comparability. 
Similar to experiments, benchmarks aim for a high control of the influencing variables and for reproducibility. On the other hand, the actual platform, tool, or technique evaluated by the benchmark can be highly variable, thus each benchmark run is similar to an experiment. However, the Empirical Standards only cover experiments with human participants.\footnote{\url{https://github.com/acmsigsoft/EmpiricalStandards/blob/master/docs/Experiments.md}}
Benchmarking is a from of controlled experimentation~\cite{tichy_ubiquity_2014}, but not yet included in the Empirical Standards.
Benchmarks require and provide particularly high levels of control.

\section{Requirements on Benchmarks}\label{s-requirements}

Benchmarking is not only relevant for research, but has also a long history in industry.
Kistowski et al.\ \cite{vKistowski2015} introduce the primary concerns of benchmark development from the perspectives of the SPEC and TPC committees, thus industrial consortia.
Benchmark candidates must undergo a process of several steps, including the definition of measurement methodologies, workload selection, and a number of rigorous benchmark acceptance tests.
Kistowski et al.\ \cite{vKistowski2015} provide a definition of the term \textit{benchmark} in the context of performance evaluation, and define a benchmark as a \textit{standard} tool for the competitive evaluation and comparison of competing systems or components according to specific characteristics, such as performance, dependability, or security.

Key characteristics of benchmarks are~\cite{vKistowski2015}:
\begin{itemize}
	\item \textbf{Relevance} How closely the benchmark behavior correlates to behaviors that are of interest to consumers of the results.
	\item \textbf{Reproducibility} The ability to consistently produce similar results when the benchmark is run with the same test configuration.
	\item \textbf{Fairness} Allowing different test configurations to compete on their merits without artificial limitations.
	\item \textbf{Verifiability} Providing confidence that a benchmark result is accurate.
	\item \textbf{Usability} Avoiding roadblocks for users to run the benchmark in their test environments.
\end{itemize}

Similar and complementary to these characteristics, Gray~\cite{Gray93} postulates relevance, portability, scalability and simplicity as basic benchmark requirements.
Essential components of benchmarks are~\cite{Sim2003}:
\begin{itemize}
	\item \textbf{Motivating Comparison} The purpose of a benchmark is to compare, and the motivation aspect refers to the need for the research area, and in turn the benchmark itself.
	\item \textbf{Task Sample} The tests in the benchmark should be a representative sample of the tasks that the method, tool or technique is expected to solve in actual practice. For performance evaluation this is a representative workload as part of the benchmark.
	\item \textbf{Performance Measures} The measurements can be made by a computer or by a human, and can be quantitative or qualitative.
\end{itemize}

Tichy~\cite{Tichy1998,tichy_ubiquity_2014} explicates that constructing benchmarks is hard work, best shared within a \textit{community}. Furthermore, benchmarks need to evolve from narrowly targeted tests to broader, generalizable tests in order to prevent overfitting for a specific goal.
Sim et al. \cite{Sim2003} further pursue the community idea and state that benchmarks must always be developed and used in the community, rather than by a single researcher.

However, it should be sufficient to start the development process of a new benchmark with a small group of researches as an offer to a larger scientific community. Such a \textit{proto-benchmark} \cite{Sim2003} can act as a template to further the discussion of the topic and to initialize the consensus process.
A proto-benchmark is a set of tests that is missing some of the above-mentioned requirements. The most common proto-benchmarks lack a performance measure~\cite{Sim2003}. Defining appropriate performance measures will be difficult in some areas of software engineering that involve human activities.

These requirements may constitute the basis for a \textit{checklist} of a benchmarking standard for empirical software engineering research. However, it is necessary to continuously scrutinize and adapt benchmarks to avoid over-optimization of research towards the benchmarks.

\section{Summary and Outlook}\label{s-conclusion}

This proposal paper does not report on an empirical research project; thus, it cannot be evaluated on the basis of the Empirical Standards. Instead, based on previous experience and literature on benchmarking in software engineering research we \textit{argue} for including benchmarks into the Empirical Standards for Software Engineering Research.

If the community (for this paper the PROPSER Workshop on Properties of Software Engineering Research) confirms my assessment that we should include benchmarking as an empirical standard, I will prepare an appropriate pull request for a benchmarking checklist at \url{https://github.com/acmsigsoft/EmpiricalStandards/}. This benchmarking checklist could constitute a separate document, or it could be an addition to the Engineering Research and/or Experiments standards. Some specific evaluation criteria for benchmarks, which are not yet included in these standards, are the following:
\begin{description}
	\item[Essential Attributes] ~
\begin{itemize}
  \item Justifies the relevance of the benchmark.
	\item Describes the experimental setup for the benchmark with sufficient detail.
	\item Specifies the synthetic workload with sufficient detail.
	\item Allows different test configurations to compete on their merits without artificial limitations.
	\item Provides confidence that a benchmark result is accurate.
	\item Avoids roadblocks for users to run the benchmark in their test environments.
	\item Provides a replication package including datasets and analytical scripts\\ (for Engineering Research this a desirable attribute, for benchmarks this is an essential attribute).
\end{itemize}
	\item[Desirable Attributes] ~
	\begin{itemize}
		\item Reports on independent replication of the benchmark.
		\item Reports on a large community that uses the benchmark.
		\item Reports on an independent organization that maintains the benchmark.
	\end{itemize}
\end{description}

\begin{acks}
I would like to thank S\"oren Henning, Lutz Prechelt, and the anonymous reviewers for their valuable feedback on earlier versions of this paper.
\end{acks}

\newpage
\balance


\begin{thebibliography}{38}


\ifx \showCODEN    \undefined \def \showCODEN     #1{\unskip}     \fi
\ifx \showDOI      \undefined \def \showDOI       #1{#1}\fi
\ifx \showISBNx    \undefined \def \showISBNx     #1{\unskip}     \fi
\ifx \showISBNxiii \undefined \def \showISBNxiii  #1{\unskip}     \fi
\ifx \showISSN     \undefined \def \showISSN      #1{\unskip}     \fi
\ifx \showLCCN     \undefined \def \showLCCN      #1{\unskip}     \fi
\ifx \shownote     \undefined \def \shownote      #1{#1}          \fi
\ifx \showarticletitle \undefined \def \showarticletitle #1{#1}   \fi
\ifx \showURL      \undefined \def \showURL       {\relax}        \fi
\providecommand\bibfield[2]{#2}
\providecommand\bibinfo[2]{#2}
\providecommand\natexlab[1]{#1}
\providecommand\showeprint[2][]{arXiv:#2}

\bibitem[\protect\citeauthoryear{Arlitt and Jin}{Arlitt and Jin}{2000}]%
        {Arlitt2000}
\bibfield{author}{\bibinfo{person}{M. Arlitt} {and} \bibinfo{person}{T. Jin}.}
  \bibinfo{year}{2000}\natexlab{}.
\newblock \showarticletitle{A workload characterization study of the 1998 World
  Cup Web site}.
\newblock \bibinfo{journal}{\emph{{IEEE} Network}} \bibinfo{volume}{14},
  \bibinfo{number}{3} (\bibinfo{year}{2000}), \bibinfo{pages}{30--37}.
\newblock
\urldef\tempurl%
\url{https://doi.org/10.1109/65.844498}
\showDOI{\tempurl}


\bibitem[\protect\citeauthoryear{Benz, Hotho, J{\"a}schke, Krause, Mitzlaff,
  Schmitz, and Stumme}{Benz et~al\mbox{.}}{2010}]%
        {Benz2010}
\bibfield{author}{\bibinfo{person}{Dominik Benz}, \bibinfo{person}{Andreas
  Hotho}, \bibinfo{person}{Robert J{\"a}schke}, \bibinfo{person}{Beate Krause},
  \bibinfo{person}{Folke Mitzlaff}, \bibinfo{person}{Christoph Schmitz}, {and}
  \bibinfo{person}{Gerd Stumme}.} \bibinfo{year}{2010}\natexlab{}.
\newblock \showarticletitle{The social bookmark and publication management
  system bibsonomy}.
\newblock \bibinfo{journal}{\emph{The {VLDB} Journal}} \bibinfo{volume}{19},
  \bibinfo{number}{6} (\bibinfo{date}{Dec.} \bibinfo{year}{2010}),
  \bibinfo{pages}{849--875}.
\newblock
\urldef\tempurl%
\url{https://doi.org/10.1007/s00778-010-0208-4}
\showDOI{\tempurl}


\bibitem[\protect\citeauthoryear{Bermbach, Wittern, and Tai}{Bermbach
  et~al\mbox{.}}{2017}]%
        {bermbach2017cloud}
\bibfield{author}{\bibinfo{person}{David Bermbach}, \bibinfo{person}{Erik
  Wittern}, {and} \bibinfo{person}{Stefan Tai}.}
  \bibinfo{year}{2017}\natexlab{}.
\newblock \bibinfo{booktitle}{\emph{Cloud service benchmarking}}.
\newblock \bibinfo{publisher}{Springer}.
\newblock


\bibitem[\protect\citeauthoryear{Bischofs, Giesecke, Gottschalk, Hasselbring,
  Warns, and Willer}{Bischofs et~al\mbox{.}}{2006}]%
        {JSS2006}
\bibfield{author}{\bibinfo{person}{L. Bischofs}, \bibinfo{person}{S. Giesecke},
  \bibinfo{person}{M. Gottschalk}, \bibinfo{person}{W. Hasselbring},
  \bibinfo{person}{T. Warns}, {and} \bibinfo{person}{S. Willer}.}
  \bibinfo{year}{2006}\natexlab{}.
\newblock \showarticletitle{Comparative evaluation of dependability
  characteristics for peer-to-peer architectural styles by simulation}.
\newblock \bibinfo{journal}{\emph{Journal of Systems and Software}}
  \bibinfo{volume}{79}, \bibinfo{number}{10} (\bibinfo{date}{Oct.}
  \bibinfo{year}{2006}), \bibinfo{pages}{1419--1432}.
\newblock
\urldef\tempurl%
\url{https://doi.org/10.1016/j.jss.2006.02.063}
\showDOI{\tempurl}


\bibitem[\protect\citeauthoryear{Gray}{Gray}{1993}]%
        {Gray93}
\bibfield{editor}{\bibinfo{person}{Jim Gray}} (Ed.).
  \bibinfo{year}{1993}\natexlab{}.
\newblock \bibinfo{booktitle}{\emph{The Benchmark Handbook for Database and
  Transaction Systems} (\bibinfo{edition}{2nd} ed.)}.
\newblock \bibinfo{publisher}{Morgan Kaufmann}.
\newblock


\bibitem[\protect\citeauthoryear{Hasselbring}{Hasselbring}{2016}]%
        {ICPE2016Keynote}
\bibfield{author}{\bibinfo{person}{Wilhelm Hasselbring}.}
  \bibinfo{year}{2016}\natexlab{}.
\newblock \showarticletitle{Microservices for Scalability: Keynote Talk
  Abstract}. In \bibinfo{booktitle}{\emph{Proceedings of the 7th ACM/SPEC
  International Conference on Performance Engineering (ICPE 2016)}} (Delft, The
  Netherlands). \bibinfo{publisher}{ACM}, \bibinfo{address}{New York, NY, USA},
  \bibinfo{pages}{133--134}.
\newblock
\showISBNx{978-1-4503-4080-9}
\urldef\tempurl%
\url{https://doi.org/10.1145/2851553.2858659}
\showDOI{\tempurl}


\bibitem[\protect\citeauthoryear{Hasselbring}{Hasselbring}{2018}]%
        {SA2018}
\bibfield{author}{\bibinfo{person}{Wilhelm Hasselbring}.}
  \bibinfo{year}{2018}\natexlab{}.
\newblock \showarticletitle{Software Architecture: Past, Present, Future}.
\newblock In \bibinfo{booktitle}{\emph{The Essence of Software Engineering}},
  \bibfield{editor}{\bibinfo{person}{Volker Gruhn} {and}
  \bibinfo{person}{R{\"u}diger Striemer}} (Eds.). \bibinfo{publisher}{Springer
  International Publishing}, \bibinfo{address}{Cham},
  \bibinfo{pages}{169--184}.
\newblock
\showISBNx{978-3-319-73897-0}
\urldef\tempurl%
\url{https://doi.org/10.1007/978-3-319-73897-0_10}
\showDOI{\tempurl}


\bibitem[\protect\citeauthoryear{Hasselbring and Steinacker}{Hasselbring and
  Steinacker}{2017}]%
        {Hasselbring2017}
\bibfield{author}{\bibinfo{person}{Wilhelm Hasselbring} {and}
  \bibinfo{person}{Guido Steinacker}.} \bibinfo{year}{2017}\natexlab{}.
\newblock \showarticletitle{Microservice Architectures for Scalability, Agility
  and Reliability in E-Commerce}. In \bibinfo{booktitle}{\emph{Proc. IEEE
  International Conference on Software Architecture Workshops}}.
\newblock
\urldef\tempurl%
\url{https://doi.org/10.1109/ICSAW.2017.11}
\showDOI{\tempurl}


\bibitem[\protect\citeauthoryear{Hasselbring and van Hoorn}{Hasselbring and van
  Hoorn}{2020}]%
        {Kieker2020}
\bibfield{author}{\bibinfo{person}{Wilhelm Hasselbring} {and}
  \bibinfo{person}{Andr{\'e} van Hoorn}.} \bibinfo{year}{2020}\natexlab{}.
\newblock \showarticletitle{Kieker: {A} monitoring framework for software
  engineering research}.
\newblock \bibinfo{journal}{\emph{Software Impacts}}  \bibinfo{volume}{5}
  (\bibinfo{date}{Aug.} \bibinfo{year}{2020}).
\newblock
\urldef\tempurl%
\url{https://doi.org/10.1016/j.simpa.2020.100019}
\showDOI{\tempurl}


\bibitem[\protect\citeauthoryear{Henning and Hasselbring}{Henning and
  Hasselbring}{2021}]%
        {BDR2021}
\bibfield{author}{\bibinfo{person}{S\"{o}ren Henning} {and}
  \bibinfo{person}{Wilhelm Hasselbring}.} \bibinfo{year}{2021}\natexlab{}.
\newblock \showarticletitle{Theodolite: Scalability Benchmarking of Distributed
  Stream Processing Engines in Microservice Architectures}.
\newblock \bibinfo{journal}{\emph{Big Data Research}} \bibinfo{volume}{25},
  \bibinfo{number}{100209} (\bibinfo{date}{July} \bibinfo{year}{2021}),
  \bibinfo{pages}{1--17}.
\newblock
\urldef\tempurl%
\url{https://doi.org/10.1016/j.bdr.2021.100209}
\showDOI{\tempurl}


\bibitem[\protect\citeauthoryear{Herbst, Kounev, and Reussner}{Herbst
  et~al\mbox{.}}{2013}]%
        {Elasticity2013}
\bibfield{author}{\bibinfo{person}{Nikolas~Roman Herbst},
  \bibinfo{person}{Samuel Kounev}, {and} \bibinfo{person}{Ralf Reussner}.}
  \bibinfo{year}{2013}\natexlab{}.
\newblock \showarticletitle{Elasticity in Cloud Computing: What It Is, and What
  It Is Not}. In \bibinfo{booktitle}{\emph{10th International Conference on
  Autonomic Computing ({ICAC} 13)}}. \bibinfo{publisher}{{USENIX} Association},
  \bibinfo{address}{San Jose, CA}, \bibinfo{pages}{23--27}.
\newblock


\bibitem[\protect\citeauthoryear{Hildebrandt, Bischofs, and
  Hasselbring}{Hildebrandt et~al\mbox{.}}{2007}]%
        {Hildebrandt2007}
\bibfield{author}{\bibinfo{person}{Dieter Hildebrandt}, \bibinfo{person}{Ludger
  Bischofs}, {and} \bibinfo{person}{Wilhelm Hasselbring}.}
  \bibinfo{year}{2007}\natexlab{}.
\newblock \showarticletitle{{RealPeer -- A Framework for Simulation-based
  Development of Peer-to-Peer Systems}}. In
  \bibinfo{booktitle}{\emph{Proceedings of the 15th Euromicro Conference on
  Parallel, Distributed and Network-based Processing (PDP 2007)}}.
  \bibinfo{publisher}{IEEE Computer Society Press}, \bibinfo{address}{Los
  Alamitos, CA, USA}, \bibinfo{pages}{490--497}.
\newblock
\urldef\tempurl%
\url{https://doi.org/10.1109/PDP.2007.70}
\showDOI{\tempurl}


\bibitem[\protect\citeauthoryear{Hildebrandt and Hasselbring}{Hildebrandt and
  Hasselbring}{2008}]%
        {RealPeer2008}
\bibfield{author}{\bibinfo{person}{Dieter Hildebrandt} {and}
  \bibinfo{person}{Wilhelm Hasselbring}.} \bibinfo{year}{2008}\natexlab{}.
\newblock \showarticletitle{Simulation-based Development of Peer-to-Peer
  Systems with the {RealPeer} Methodology and Framework}.
\newblock \bibinfo{journal}{\emph{Journal of Systems Architecture}}
  \bibinfo{volume}{54}, \bibinfo{number}{9} (\bibinfo{date}{Sept.}
  \bibinfo{year}{2008}), \bibinfo{pages}{849--860}.
\newblock
\urldef\tempurl%
\url{https://doi.org/10.1016/j.sysarc.2008.01.010}
\showDOI{\tempurl}


\bibitem[\protect\citeauthoryear{Johanson and Hasselbring}{Johanson and
  Hasselbring}{2017}]%
        {ESE2017}
\bibfield{author}{\bibinfo{person}{Arne Johanson} {and}
  \bibinfo{person}{Wilhelm Hasselbring}.} \bibinfo{year}{2017}\natexlab{}.
\newblock \showarticletitle{Effectiveness and efficiency of a domain-specific
  language for high-performance marine ecosystem simulation: a controlled
  experiment}.
\newblock \bibinfo{journal}{\emph{Empirical Software Engineering}}
  \bibinfo{volume}{22}, \bibinfo{number}{4} (\bibinfo{date}{Aug.}
  \bibinfo{year}{2017}), \bibinfo{pages}{2206--2236}.
\newblock
\showISSN{1382-3256}
\urldef\tempurl%
\url{https://doi.org/10.1007/s10664-016-9483-z}
\showDOI{\tempurl}


\bibitem[\protect\citeauthoryear{Johanson and Hasselbring}{Johanson and
  Hasselbring}{2018}]%
        {CiSE2018}
\bibfield{author}{\bibinfo{person}{Arne Johanson} {and} \bibinfo{person}{Wilhem
  Hasselbring}.} \bibinfo{year}{2018}\natexlab{}.
\newblock \showarticletitle{Software Engineering for Computational Science:
  Past, Present, Future}.
\newblock \bibinfo{journal}{\emph{Computing in Science {\&} Engineering}}
  \bibinfo{volume}{20}, \bibinfo{number}{2} (\bibinfo{date}{March}
  \bibinfo{year}{2018}), \bibinfo{pages}{90--109}.
\newblock
\urldef\tempurl%
\url{https://doi.org/10.1109/MCSE.2018.021651343}
\showDOI{\tempurl}


\bibitem[\protect\citeauthoryear{Johanson, Oschlies, Hasselbring, and
  Worm}{Johanson et~al\mbox{.}}{2017}]%
        {Johanson:2017}
\bibfield{author}{\bibinfo{person}{Arne Johanson}, \bibinfo{person}{Andreas
  Oschlies}, \bibinfo{person}{Wilhelm Hasselbring}, {and}
  \bibinfo{person}{Boris Worm}.} \bibinfo{year}{2017}\natexlab{}.
\newblock \showarticletitle{{SPRAT: A spatially-explicit marine ecosystem model
  based on population balance equations}}.
\newblock \bibinfo{journal}{\emph{Ecological Modelling}}  \bibinfo{volume}{349}
  (\bibinfo{date}{April} \bibinfo{year}{2017}), \bibinfo{pages}{11--25}.
\newblock
\urldef\tempurl%
\url{https://doi.org/10.1016/j.ecolmodel.2017.01.020}
\showDOI{\tempurl}


\bibitem[\protect\citeauthoryear{Karimov, Rabl, Katsifodimos, Samarev,
  Heiskanen, and Markl}{Karimov et~al\mbox{.}}{2018}]%
        {Karimov2018}
\bibfield{author}{\bibinfo{person}{Jeyhun Karimov}, \bibinfo{person}{Tilmann
  Rabl}, \bibinfo{person}{Asterios Katsifodimos}, \bibinfo{person}{Roman
  Samarev}, \bibinfo{person}{Henri Heiskanen}, {and} \bibinfo{person}{Volker
  Markl}.} \bibinfo{year}{2018}\natexlab{}.
\newblock \showarticletitle{Benchmarking Distributed Stream Data Processing
  Systems}. In \bibinfo{booktitle}{\emph{IEEE 34th International Conference on
  Data Engineering (ICDE)}}. \bibinfo{pages}{1507--1518}.
\newblock
\urldef\tempurl%
\url{https://doi.org/10.1109/ICDE.2018.00169}
\showDOI{\tempurl}


\bibitem[\protect\citeauthoryear{Knoche and Eichelberger}{Knoche and
  Eichelberger}{2018}]%
        {Knoche2018}
\bibfield{author}{\bibinfo{person}{Holger Knoche} {and} \bibinfo{person}{Holger
  Eichelberger}.} \bibinfo{year}{2018}\natexlab{}.
\newblock \showarticletitle{Using the Raspberry Pi and Docker for Replicable
  Performance Experiments}. In \bibinfo{booktitle}{\emph{Proceedings of the
  2018 {ACM}/{SPEC} International Conference on Performance Engineering}}.
  \bibinfo{publisher}{{ACM}}.
\newblock
\urldef\tempurl%
\url{https://doi.org/10.1145/3184407.3184431}
\showDOI{\tempurl}


\bibitem[\protect\citeauthoryear{Knoche and Hasselbring}{Knoche and
  Hasselbring}{2019}]%
        {EMISA2019}
\bibfield{author}{\bibinfo{person}{Holger Knoche} {and}
  \bibinfo{person}{Wilhelm Hasselbring}.} \bibinfo{year}{2019}\natexlab{}.
\newblock \showarticletitle{Drivers and Barriers for Microservice Adoption -- A
  Survey among Professionals in {Germany}}.
\newblock \bibinfo{journal}{\emph{Enterprise Modelling and Information Systems
  Architectures (EMISAJ) -- International Journal of Conceptual Modeling}}
  \bibinfo{volume}{14}, \bibinfo{number}{1} (\bibinfo{year}{2019}),
  \bibinfo{pages}{1--35}.
\newblock
\urldef\tempurl%
\url{https://doi.org/10.18417/emisa.14.1}
\showDOI{\tempurl}


\bibitem[\protect\citeauthoryear{Kounev, Lange, and von Kistowski}{Kounev
  et~al\mbox{.}}{2020}]%
        {kounev2020systems}
\bibfield{author}{\bibinfo{person}{Samuel Kounev},
  \bibinfo{person}{Klaus-Dieter Lange}, {and} \bibinfo{person}{J{\'o}akim von
  Kistowski}.} \bibinfo{year}{2020}\natexlab{}.
\newblock \bibinfo{booktitle}{\emph{Systems Benchmarking for Scientists and
  Engineers}}.
\newblock \bibinfo{publisher}{Springer}.
\newblock


\bibitem[\protect\citeauthoryear{Nanni, Mitra, Magnusson, and Dietz}{Nanni
  et~al\mbox{.}}{2017}]%
        {Nanni2017}
\bibfield{author}{\bibinfo{person}{Federico Nanni}, \bibinfo{person}{Bhaskar
  Mitra}, \bibinfo{person}{Matt Magnusson}, {and} \bibinfo{person}{Laura
  Dietz}.} \bibinfo{year}{2017}\natexlab{}.
\newblock \showarticletitle{Benchmark for Complex Answer Retrieval}. In
  \bibinfo{booktitle}{\emph{Proceedings of the {ACM} {SIGIR} International
  Conference on Theory of Information Retrieval}}. \bibinfo{publisher}{{ACM}}.
\newblock
\urldef\tempurl%
\url{https://doi.org/10.1145/3121050.3121099}
\showDOI{\tempurl}


\bibitem[\protect\citeauthoryear{Ralph}{Ralph}{2021}]%
        {Empirical2021}
\bibfield{author}{\bibinfo{person}{Paul Ralph}.}
  \bibinfo{year}{2021}\natexlab{}.
\newblock \showarticletitle{{ACM} {SIGSOFT} Empirical Standards Released}.
\newblock \bibinfo{journal}{\emph{{ACM} {SIGSOFT} Software Engineering Notes}}
  \bibinfo{volume}{46}, \bibinfo{number}{1} (\bibinfo{date}{Feb.}
  \bibinfo{year}{2021}), \bibinfo{pages}{19--19}.
\newblock
\urldef\tempurl%
\url{https://doi.org/10.1145/3437479.3437483}
\showDOI{\tempurl}


\bibitem[\protect\citeauthoryear{Ralph, bin Ali, Baltes, Bianculli, Diaz,
  Dittrich, Ernst, Felderer, Feldt, Filieri, de~França, Furia, Gay, Gold,
  Graziotin, He, Hoda, Juristo, Kitchenham, Lenarduzzi, Martínez, Melegati,
  Mendez, Menzies, Molleri, Pfahl, Robbes, Russo, Saarimäki, Sarro, Taibi,
  Siegmund, Spinellis, Staron, Stol, Storey, Taibi, Tamburri, Torchiano,
  Treude, Turhan, Wang, and Vegas}{Ralph et~al\mbox{.}}{2021}]%
        {ralph2021acm}
\bibfield{author}{\bibinfo{person}{Paul Ralph}, \bibinfo{person}{Nauman bin
  Ali}, \bibinfo{person}{Sebastian Baltes}, \bibinfo{person}{Domenico
  Bianculli}, \bibinfo{person}{Jessica Diaz}, \bibinfo{person}{Yvonne
  Dittrich}, \bibinfo{person}{Neil Ernst}, \bibinfo{person}{Michael Felderer},
  \bibinfo{person}{Robert Feldt}, \bibinfo{person}{Antonio Filieri},
  \bibinfo{person}{Breno Bernard~Nicolau de França},
  \bibinfo{person}{Carlo~Alberto Furia}, \bibinfo{person}{Greg Gay},
  \bibinfo{person}{Nicolas Gold}, \bibinfo{person}{Daniel Graziotin},
  \bibinfo{person}{Pinjia He}, \bibinfo{person}{Rashina Hoda},
  \bibinfo{person}{Natalia Juristo}, \bibinfo{person}{Barbara Kitchenham},
  \bibinfo{person}{Valentina Lenarduzzi}, \bibinfo{person}{Jorge Martínez},
  \bibinfo{person}{Jorge Melegati}, \bibinfo{person}{Daniel Mendez},
  \bibinfo{person}{Tim Menzies}, \bibinfo{person}{Jefferson Molleri},
  \bibinfo{person}{Dietmar Pfahl}, \bibinfo{person}{Romain Robbes},
  \bibinfo{person}{Daniel Russo}, \bibinfo{person}{Nyyti Saarimäki},
  \bibinfo{person}{Federica Sarro}, \bibinfo{person}{Davide Taibi},
  \bibinfo{person}{Janet Siegmund}, \bibinfo{person}{Diomidis Spinellis},
  \bibinfo{person}{Miroslaw Staron}, \bibinfo{person}{Klaas Stol},
  \bibinfo{person}{Margaret-Anne Storey}, \bibinfo{person}{Davide Taibi},
  \bibinfo{person}{Damian Tamburri}, \bibinfo{person}{Marco Torchiano},
  \bibinfo{person}{Christoph Treude}, \bibinfo{person}{Burak Turhan},
  \bibinfo{person}{Xiaofeng Wang}, {and} \bibinfo{person}{Sira Vegas}.}
  \bibinfo{year}{2021}\natexlab{}.
\newblock \bibinfo{title}{Empirical Standards for Software Engineering
  Research}.
\newblock
\newblock
\urldef\tempurl%
\url{http://arxiv.org/abs/2010.03525}
\showURL{%
\tempurl}
\newblock
\shownote{Version 0.2.0.}


\bibitem[\protect\citeauthoryear{Sim, Easterbrook, and Holt}{Sim
  et~al\mbox{.}}{2003}]%
        {Sim2003}
\bibfield{author}{\bibinfo{person}{Susan~Elliott Sim}, \bibinfo{person}{Steve
  Easterbrook}, {and} \bibinfo{person}{Richard~C. Holt}.}
  \bibinfo{year}{2003}\natexlab{}.
\newblock \showarticletitle{Using benchmarking to advance research: a challenge
  to software engineering}. In \bibinfo{booktitle}{\emph{25th International
  Conference on Software Engineering}}. \bibinfo{publisher}{{IEEE}}.
\newblock
\urldef\tempurl%
\url{https://doi.org/10.1109/icse.2003.1201189}
\showDOI{\tempurl}


\bibitem[\protect\citeauthoryear{Smith and Williams}{Smith and
  Williams}{2002}]%
        {smith2002performance}
\bibfield{author}{\bibinfo{person}{Connie~U Smith} {and}
  \bibinfo{person}{Lloyd~G Williams}.} \bibinfo{year}{2002}\natexlab{}.
\newblock \bibinfo{booktitle}{\emph{Performance solutions: a practical guide to
  creating responsive, scalable software}}.
\newblock \bibinfo{publisher}{Addison-Wesley}.
\newblock


\bibitem[\protect\citeauthoryear{Tichy}{Tichy}{1998}]%
        {Tichy1998}
\bibfield{author}{\bibinfo{person}{Walter~F. Tichy}.}
  \bibinfo{year}{1998}\natexlab{}.
\newblock \showarticletitle{Should computer scientists experiment more?}
\newblock \bibinfo{journal}{\emph{Computer}} \bibinfo{volume}{31},
  \bibinfo{number}{5} (\bibinfo{date}{May} \bibinfo{year}{1998}),
  \bibinfo{pages}{32--40}.
\newblock
\urldef\tempurl%
\url{https://doi.org/10.1109/2.675631}
\showDOI{\tempurl}


\bibitem[\protect\citeauthoryear{Tichy}{Tichy}{2014}]%
        {tichy_ubiquity_2014}
\bibfield{author}{\bibinfo{person}{Walter~F. Tichy}.}
  \bibinfo{year}{2014}\natexlab{}.
\newblock \showarticletitle{Where's the Science in Software Engineering?
  Ubiquity Symposium: The Science in Computer Science}.
\newblock \bibinfo{journal}{\emph{Ubiquity}}  \bibinfo{volume}{2014}
  (\bibinfo{date}{March} \bibinfo{year}{2014}), \bibinfo{pages}{1--6}.
\newblock
\urldef\tempurl%
\url{https://doi.org/10.1145/2590528.2590529}
\showDOI{\tempurl}


\bibitem[\protect\citeauthoryear{v.~Kistowski, Arnold, Huppler, Lange, Henning,
  and Cao}{v.~Kistowski et~al\mbox{.}}{2015}]%
        {vKistowski2015}
\bibfield{author}{\bibinfo{person}{J{\'{o}}akim v. Kistowski},
  \bibinfo{person}{Jeremy~A. Arnold}, \bibinfo{person}{Karl Huppler},
  \bibinfo{person}{Klaus-Dieter Lange}, \bibinfo{person}{John~L. Henning},
  {and} \bibinfo{person}{Paul Cao}.} \bibinfo{year}{2015}\natexlab{}.
\newblock \showarticletitle{How to Build a Benchmark}. In
  \bibinfo{booktitle}{\emph{Proceedings of the 6th {ACM}/{SPEC} International
  Conference on Performance Engineering}}. \bibinfo{publisher}{{ACM}}.
\newblock
\urldef\tempurl%
\url{https://doi.org/10.1145/2668930.2688819}
\showDOI{\tempurl}


\bibitem[\protect\citeauthoryear{van Hoorn, Rohr, and Hasselbring}{van Hoorn
  et~al\mbox{.}}{2008}]%
        {vanHoorn2008}
\bibfield{author}{\bibinfo{person}{Andr{\'{e}} van Hoorn},
  \bibinfo{person}{Matthias Rohr}, {and} \bibinfo{person}{Wilhelm
  Hasselbring}.} \bibinfo{year}{2008}\natexlab{}.
\newblock \showarticletitle{Generating Probabilistic and Intensity-Varying
  Workload for Web-Based Software Systems}.
\newblock In \bibinfo{booktitle}{\emph{Performance Evaluation: Metrics, Models
  and Benchmarks}}. \bibinfo{publisher}{Springer}, \bibinfo{pages}{124--143}.
\newblock
\urldef\tempurl%
\url{https://doi.org/10.1007/978-3-540-69814-2_9}
\showDOI{\tempurl}


\bibitem[\protect\citeauthoryear{van Hoorn, V{\"o}gele, Schulz, Hasselbring,
  and Krcmar}{van Hoorn et~al\mbox{.}}{2014}]%
        {ValueTools2014}
\bibfield{author}{\bibinfo{person}{Andre van Hoorn}, \bibinfo{person}{Christian
  V{\"o}gele}, \bibinfo{person}{Eike Schulz}, \bibinfo{person}{Wilhelm
  Hasselbring}, {and} \bibinfo{person}{Helmut Krcmar}.}
  \bibinfo{year}{2014}\natexlab{}.
\newblock \showarticletitle{Automatic Extraction of Probabilistic Workload
  Specifications for Load Testing Session-Based Application Systems}. In
  \bibinfo{booktitle}{\emph{Proceedings of the 8th International Conference on
  Performance Evaluation Methodologies and Tools (ValueTools 2014)}}
  (Bratislava, Slovakia). \bibinfo{publisher}{ICST}, \bibinfo{pages}{139--146}.
\newblock
\showISBNx{978-1-63190-057-0}
\urldef\tempurl%
\url{https://doi.org/10.4108/icst.valuetools.2014.258171}
\showDOI{\tempurl}


\bibitem[\protect\citeauthoryear{van Hoorn, Waller, and Hasselbring}{van Hoorn
  et~al\mbox{.}}{2012}]%
        {KiekerICPE2012}
\bibfield{author}{\bibinfo{person}{Andr{\'{e}} van Hoorn}, \bibinfo{person}{Jan
  Waller}, {and} \bibinfo{person}{Wilhelm Hasselbring}.}
  \bibinfo{year}{2012}\natexlab{}.
\newblock \showarticletitle{Kieker: {A} Framework for Application Performance
  Monitoring and Dynamic Software Analysis}. In
  \bibinfo{booktitle}{\emph{Proceedings of the 3rd ACM/SPEC International
  Conference on Performance Engineering (ICPE 2012)}} (Boston, Massachusetts,
  USA, April 22-25, 2012). \bibinfo{publisher}{ACM}, \bibinfo{pages}{247--248}.
\newblock
\urldef\tempurl%
\url{https://doi.org/10.1145/2188286.2188326}
\showDOI{\tempurl}


\bibitem[\protect\citeauthoryear{Vieira, Madeira, Sachs, and Kounev}{Vieira
  et~al\mbox{.}}{2012}]%
        {Vieira2012}
\bibfield{author}{\bibinfo{person}{Marco Vieira}, \bibinfo{person}{Henrique
  Madeira}, \bibinfo{person}{Kai Sachs}, {and} \bibinfo{person}{Samuel
  Kounev}.} \bibinfo{year}{2012}\natexlab{}.
\newblock \showarticletitle{Resilience Benchmarking}.
\newblock In \bibinfo{booktitle}{\emph{Resilience Assessment and Evaluation of
  Computing Systems}}. \bibinfo{publisher}{Springer},
  \bibinfo{pages}{283--301}.
\newblock
\urldef\tempurl%
\url{https://doi.org/10.1007/978-3-642-29032-9_14}
\showDOI{\tempurl}


\bibitem[\protect\citeauthoryear{V{\"o}gele, van Hoorn, Schulz, Hasselbring,
  and Krcmar}{V{\"o}gele et~al\mbox{.}}{2018}]%
        {WESSBAS2018}
\bibfield{author}{\bibinfo{person}{Christian V{\"o}gele},
  \bibinfo{person}{Andr{\'e} van Hoorn}, \bibinfo{person}{Eike Schulz},
  \bibinfo{person}{Wilhelm Hasselbring}, {and} \bibinfo{person}{Helmut
  Krcmar}.} \bibinfo{year}{2018}\natexlab{}.
\newblock \showarticletitle{{WESSBAS:} extraction of probabilistic workload
  specifications for load testing and performance prediction---a model-driven
  approach for session-based application systems}.
\newblock \bibinfo{journal}{\emph{Software {\&} Systems Modeling}}
  \bibinfo{volume}{17}, \bibinfo{number}{2} (\bibinfo{date}{May}
  \bibinfo{year}{2018}), \bibinfo{pages}{443--477}.
\newblock
\urldef\tempurl%
\url{https://doi.org/10.1007/s10270-016-0566-5}
\showDOI{\tempurl}


\bibitem[\protect\citeauthoryear{von Kistowski, Eismann, Schmitt, Bauer,
  Grohmann, and Kounev}{von Kistowski et~al\mbox{.}}{2018}]%
        {vonKistowski2018}
\bibfield{author}{\bibinfo{person}{Joakim von Kistowski},
  \bibinfo{person}{Simon Eismann}, \bibinfo{person}{Norbert Schmitt},
  \bibinfo{person}{Andre Bauer}, \bibinfo{person}{Johannes Grohmann}, {and}
  \bibinfo{person}{Samuel Kounev}.} \bibinfo{year}{2018}\natexlab{}.
\newblock \showarticletitle{{TeaStore}: A Micro-Service Reference Application
  for Benchmarking, Modeling and Resource Management Research}. In
  \bibinfo{booktitle}{\emph{2018 {IEEE} 26th International Symposium on
  Modeling, Analysis, and Simulation of Computer and Telecommunication Systems
  ({MASCOTS})}}. \bibinfo{publisher}{{IEEE}}, \bibinfo{pages}{223--236}.
\newblock
\urldef\tempurl%
\url{https://doi.org/10.1109/mascots.2018.00030}
\showDOI{\tempurl}


\bibitem[\protect\citeauthoryear{von Massow, van Hoorn, and Hasselbring}{von
  Massow et~al\mbox{.}}{2011}]%
        {ECSA2011Massow}
\bibfield{author}{\bibinfo{person}{Robert von Massow}, \bibinfo{person}{Andr\'e
  van Hoorn}, {and} \bibinfo{person}{Wilhelm Hasselbring}.}
  \bibinfo{year}{2011}\natexlab{}.
\newblock \showarticletitle{Performance Simulation of Runtime Reconfigurable
  Component-Based Software Architectures}. In
  \bibinfo{booktitle}{\emph{Software Architecture (Proceedings ECSA 2011)}}
  \emph{(\bibinfo{series}{Lecture Notes in Computer Science},
  Vol.~\bibinfo{volume}{6903})}, \bibfield{editor}{\bibinfo{person}{Ivica
  Crnkovic}, \bibinfo{person}{Volker Gruhn}, {and} \bibinfo{person}{Matthias
  Book}} (Eds.). \bibinfo{publisher}{Springer-Verlag}, \bibinfo{pages}{43--58}.
\newblock
\showISBNx{978-3-642-23797-3}
\urldef\tempurl%
\url{https://doi.org/10.1007/978-3-642-23798-0_5}
\showDOI{\tempurl}


\bibitem[\protect\citeauthoryear{Waller, Ehmke, and Hasselbring}{Waller
  et~al\mbox{.}}{2015}]%
        {SEN2015}
\bibfield{author}{\bibinfo{person}{Jan Waller}, \bibinfo{person}{Nils~C.
  Ehmke}, {and} \bibinfo{person}{Wilhelm Hasselbring}.}
  \bibinfo{year}{2015}\natexlab{}.
\newblock \showarticletitle{Including Performance Benchmarks into Continuous
  Integration to Enable {DevOps}}.
\newblock \bibinfo{journal}{\emph{SIGSOFT Softw. Eng. Notes}}
  \bibinfo{volume}{40}, \bibinfo{number}{2} (\bibinfo{date}{March}
  \bibinfo{year}{2015}), \bibinfo{pages}{1--4}.
\newblock
\urldef\tempurl%
\url{https://doi.org/10.1145/2735399.2735416}
\showDOI{\tempurl}


\bibitem[\protect\citeauthoryear{Waller and Hasselbring}{Waller and
  Hasselbring}{2012}]%
        {MSEPT2012}
\bibfield{author}{\bibinfo{person}{Jan Waller} {and} \bibinfo{person}{Wilhelm
  Hasselbring}.} \bibinfo{year}{2012}\natexlab{}.
\newblock \showarticletitle{A Comparison of the Influence of Different
  Multi-Core Processors on the Runtime Overhead for Application-Level
  Monitoring}. In \bibinfo{booktitle}{\emph{Proceedings of the International
  Conference on Multicore Software Engineering, Performance, and Tools (MSEPT
  2012)}} \emph{(\bibinfo{series}{Lecture Notes in Computer Science},
  Vol.~\bibinfo{volume}{7303})}. \bibinfo{publisher}{Springer Berlin /
  Heidelberg}, \bibinfo{pages}{42--53}.
\newblock
\urldef\tempurl%
\url{https://doi.org/10.1007/978-3-642-31202-1_5}
\showDOI{\tempurl}


\bibitem[\protect\citeauthoryear{Wohlin}{Wohlin}{2021}]%
        {Wohlin2021}
\bibfield{author}{\bibinfo{person}{Claes Wohlin}.}
  \bibinfo{year}{2021}\natexlab{}.
\newblock \showarticletitle{Case Study Research in Software Engineering---It is
  a Case, and it is a Study, but is it a Case Study?}
\newblock \bibinfo{journal}{\emph{Information and Software Technology}}
  \bibinfo{volume}{133} (\bibinfo{date}{May} \bibinfo{year}{2021}).
\newblock
\urldef\tempurl%
\url{https://doi.org/10.1016/j.infsof.2021.106514}
\showDOI{\tempurl}


\end{thebibliography}
\end{document}